\def\conv{\alpha}

\def\eg{{\it e.g.}}
\def\ie{{\it i.e.}}

\def\fields{{\cal F}}
\def\theories{{\cal Q}}
\def\redef{{\cal G}}

\def\gmn{{g_{\mu\nu}}}

\null

\centerline{\bf The Renormalization Group, Systems of Units and the Hierarchy Problem}
\centerline{Roberto Percacci}
\centerline{SISSA, via Beirut 4, I-34014 Trieste, Italy}
\centerline{INFN, Sezione di Trieste, Italy}

\midinsert
\narrower
In the context of the Renormalization Group (RG) for gravity
I discuss the role of field rescalings and their relation to choices of units.
I concentrate on a simple Higgs model coupled to gravity, 
where natural choices of units can be based on Newton's constant 
or on the Higgs mass. These quantities are not invariant under the RG, 
and the ratio between the units is scale-dependent.
In the toy model, strong RG running occurs in the intermediate 
regime between the Higgs and the Planck scale, 
reproducing the results of the Randall-Sundrum I model.
Possible connections with the problem of the mass hierarchy are pointed out.

\endinsert


\bigskip

\leftline{\bf 1. Introduction}
\smallskip
In a Wilsonian approach to the Renormalization Group (RG), physics
at some energy scale $k$ is described by an effective action $\Gamma_k$
which can be regarded as the result of having functionally integrated
all field fluctuations with momenta larger than $k$ [1].
Thus $k$ can be interpreted as an infrared cutoff and
the RG describes the dependence of the action on this cutoff.
An infinitesimal RG transformation is basically an integration 
over fluctuations of the fields with momenta ranging from $k$ to $k-dk$.
A sequence of infinitesimal RG transformations defines a flow in the
space of all possible actions. I will call this the ``basic'' RG flow.

In Quantum Field Theory (QFT), one can use the freedom of redefining the 
fundamental fields to eliminate some couplings from the action.
The most familiar example, and the one that we will be dealing with 
in this paper, is normalizing the coefficient of the kinetic term
by means of a global rescaling of the fields.
If one starts from such a normalized action and performs a RG transformation, 
the resulting action will generally not be normalized anymore, 
but it can be brought back to normalized form by means of a field redefinition.
It is therefore customary to include in the definition of a RG transformation
also this field redefinition.
It is also convenient (and for numerical simulations even necessary)
to work with dimensionless variables;
this can be achieved by multiplying every coupling by a suitable power
of $k$, \ie\ using $k$ as unit of mass.
(Similarly, in QFT on a lattice one usually takes the lattice spacing as unit of length.)
The RG transformation changes the value of
the cutoff, and to compensate this it is customary 
to include in the definition of a RG transformation also a rescaling
of lengths and momenta.

Thus, the textbook definition of a ``complete'' RG transformation in QFT or
statistical mechanics involves three steps:
(i) functional integration over field fluctuations, (ii) rescaling
of the field, (iii) rescaling of coordinates and momenta.
This defines a flow in the space of normalized actions.
In this paper I discuss the use of field rescalings in the presence of gravity.
Because of the special geometrical status of the metric field,
its rescalings can be interpreted as changes in the unit of length.
Therefore, when the metric is dynamical, step (iii) is just a special
case of step (ii), applied to the metric.
Furthermore, normalizing the metric is equivalent to a choice of units.
Different choices of units lead to different RG flows and,
as a consequence of the rescalings, the normalized metric becomes scale--dependent.
This discussion will be used to clarify some points related to the
RG flow in the theory of gravity.
In particular, it will be shown that earlier results on the RG flow
in Einstein's theory will not be affected by consideration of these
rescalings.

In the second part of the paper I will use these insights to 
show how certain five--dimensional models
can be seen as geometrical reformulations of the four--dimensional RG flow.
In order to make the discussion more concrete, we will analyze in some detail 
a toy model consisting of gravity coupled to a scalar field.
Assuming that this model reflects some features of the real world,
the scalar field will be referred to as the ``Higgs field'' 
and its physical mass will be assumed to be in the TeV range.
In this model, various choices of units are possible
\footnote{$^*$}{As is customary in QFT, we take $\hbar$ as unit of action 
and $c$ as unit of velocity. 
With these choices, everything has the dimension of a power of length (or mass).
A system of units is then defined by choosing a standard of length (or mass).}:
cutoff units, Planck units based on Newton's constant and Higgs units
based on the mass of the scalar field, each leading to a different
form of the RG flow.
The conversion ratio between Higgs and Planck units is
$$
\conv={m_H\over m_P}=m_H\sqrt{G}\approx 10^{-16}\ . \eqno (1)
$$

Insofar as these units are built with fundamental physical variables,
the appearance of such a small ratio is puzzling 
\footnote{$^{\ddag}$}{see [2] for recent discussions
about fundamental constants and fundamental units}.
This was noted already long ago by Dirac, who considered various very
large numbers occurring in Nature [3]. 
One such number was the ratio of the gravitational to electromagnetic 
force between an electron and a proton.
Dirac's response was to assume that a very large number can
only arise as a result of another number being very large.
Another very large number which turns out to be of comparable
magnitude is the age of the universe, $\tau$, expressed in atomic units.
Dirac postulated that this rough coincidence is due to some as yet undiscovered law,
and that to maintain it in the course of time Newton's constant, 
when measured in atomic units, would decrease as $\tau^{-1}$.
Experimental bounds are now strong enough to rule out this behaviour
(for a review of observations see e.g. [4]).

The mass of the electron and
the mass of the proton have quite different origins, and for our discussion
it will be convenient to replace Dirac's original electron and proton 
by any pair of charged pointlike particles (quarks or leptons).
The mass of all these particles is believed to originate from their
Yukawa interaction with the Higgs VEV; likewise, the Higgs mass $m_H$ 
originates from the interaction of the scalar quanta with the VEV.
Thus the ratio (1) is a good representative of the problem posed by
the smallness of the mass of any known pointlike particle, in Planck units.
It is also closely related to Dirac's problem, because the ratio of 
gravitational to electric forces for pairs of charged pointlike particles,
up to dimensionless couplings, is essentially the square of (1). 

In particle physics the smallness of (1) is known as the hierarchy problem.
In perturbative QFT, $m_H^2$ can be thought of as the sum of a classical (``bare'')
value and a quantum correction. Since the dominant graph for the radiative correction of $m_H^2$
is quadratically divergent, the quantum correction is proportional to the square of the
UV cutoff, which is presumably of the order of Planck's mass.
In order for the observed value to be so much smaller than the Planck mass, 
the bare value should cancel the quantum correction with an extraordinarily high precision. 
While not necessarily ``wrong'', this fine tuning is very artificial and should
probably be taken as a hint that the perturbative approach 
does not describe properly what is happening.

Some years ago Randall and Sundrum (RS) proposed an elegant approach to the
hierarchy problem, based on a classical five-dimensional model [5].
They start from the five-dimensional dynamics of gravity coupled to two parallel three-branes.
It is assumed that matter fields are somehow confined to one of the branes,
which is to be identified with our four-dimensional world,
while gravity propagates in the five dimensional bulk.
This configuration, with a five-dimensional Anti-de Sitter (AdS) metric between the
branes, can be made to solve the equations of motion provided the
cosmological constants in the bulk and in the branes are properly matched.
It then emerges that any mass parameter in the ``physical'' brane, 
measured relative to the four-dimensional metric which solves Einstein's equations,
is suppressed by an exponential factor with respect to the fundamental parameters
appearing in the action.

Returning now to the RG flow, we will see that the equations 
for the running parameters of the toy model in certain approximations
are identical to the ones found by RS in [5]. With hindsight, this is not surprising:
based on ideas that originate from the AdS/CFT correspondence [6],
the RS model has been given a ``holographic'' interpretation, 
whereby the fifth coordinate in the Anti-de Sitter (AdS) space can be regarded 
as the logarithm of a RG scale [7].
The five-dimensional slice of AdS space that is used in the RS model
reproduces precisely the leading scaling behaviour of the
flat metric when the RG equations are written in cutoff or Higgs units.
There is however a fundamental difference in interpretation.
The RG equations used in this paper are not derived from the classical 
higher--dimensional Einstein equations:
instead, they are calculated from first principles in the four-dimensional QFT
and the five-dimensional AdS metric is merely a suggestive auxiliary construction.
In fact, the dynamics of gravity plays a relatively little role in this discussion:
the contribution of gravitons to the beta functions is suppressed in the regime that 
is of interest for the hierarchy problem, and the main role of gravity is to set the
initial conditions for the RG evolution.
Nevertheless, it will be clear that the natural framework for this discussion
is in the presence of dynamical gravity.

This paper is organized as follows.
Section 2 discusses the invariances of the action under scalings of the fields,
the notion of ``redundant'' or ``inessential'' couplings 
and some peculiarities of the RG applied to gravity. 
In Section 3 these notions are illustrated in a toy model of gravity coupled to a scalar field;
various choices of units are discussed, each leading to a different form of the RG equations.
Section 4 contains the actual behaviour of the ratio $\conv$ as a function 
of $k$ in the toy model.
In Section 5 the RS model is reconstructed starting from the four--dimensional RG flow;
some general consequences of the running of the units are illustrated.
Finally in Section 6 I discuss the use of dimensionful quantities in Quantum Gravity,
possible connections of the RG flow to the hierarchy problem and
I present some final comments and conclusions.

\medskip
\leftline{\bf 2. Field rescalings and redundant couplings}
\smallskip
Let us assume that physics at a certain energy scale $k$ is described, 
to the desired degree of accuracy, by an effective action $\Gamma_k$,
a functional depending on certain fields $\phi_A$ and coupling constants $g_i$ 
\footnote{$^\dagger$}{The word ``effective'' 
is used here in the same sense as in ``effective field theory''.
}.
The action describing the physics at some lower energy scale $k-dk$,
is given by a functional integral over all fluctuations
of the fields with momenta in the range $k-dk<|q|<k$, using $\Gamma_k$
as the classical action.
In general, even if $\Gamma_k$ contained only a finite number of terms,
this functional integral will produce an effective action $\Gamma_{k-dk}$
with infinitely many effective couplings.
Therefore, $\Gamma_k$ should be thought of from the outset
as ``the most general action'' for the fields.
It will have the form
$$
\Gamma_k(g_i,\phi_A)=\sum_i g_i(k) {\cal O}_i(\phi_A)\ ,\eqno(2)
$$
where ${\cal O}_i$ are all operators constructed with the fields
and their derivatives, which are compatible with the symmetries of the theory
and belong to a given class of functionals (\eg\ one may or may not require locality).
The dependence of the effective action on $k$ is given by
$$
\partial_t\Gamma_k(g_i,\phi_A)=\sum_i \beta_i(g_j,k) {\cal O}_i(\phi_A)\ ,\eqno(3)
$$
where $t=\log(k/k_0)$ (for some arbitrary initial value $k_0$) and
$\beta_i(g_j,k)=\partial_t g_i$ are the ``basic'' beta functions,
describing the change of the effective action when one integrates out
fluctuations of the fields with momenta between $k$ and $k-dk$.
Note that since only an infinitesimal range of momenta is involved, 
the beta functions defined in this way are finite, 
independently of the UV behaviour of the theory.

Once the beta functions in (3) are known, one can start from any arbitrary initial point 
and follow the RG trajectory in either direction.
It happens very often that the flow cannot be integrated towards the UV beyond
a certain limiting scale $\Lambda$. 
In this case the theory can only be used for $k<\Lambda$;
it is called an ``effective theory'' or a ``cutoff theory''
and $\Lambda$ is the physical cutoff (as opposed to a mathematical cutoff,
which is used to regulate divergences,
or to $k$, which is an artificial device 
used to compute the scale dependence of the action).
It may happen, however, that the limit $t\to\infty$
can be taken; in this case the theory is said to be ``fundamental''.
A fundamental QFT is a self-consistent description of a certain set
of physical phenomena which is valid up to arbitrarily high energy scales
and does not need to refer to anything else outside it.

We can think of the space of all theories as an
infinite dimensional manifold $\theories$, whose coordinates are the
coupling constants $g_i$ appearing in the action.
Similarly, let us denote $\fields$ the infinite dimensional space
of all field configurations, whose coordinates are the fields $\phi_A(x)$.
We can think of $\Gamma_k(\phi_A,g_i)$, for fixed $k$, as a functional on 
$\fields\times\theories$.
This description generally contains more parameters than are strictly necessary 
to describe the physics.
The fields $\phi_A$ are integration variables, and a redefinition 
of the fields does not change the physical content of the theory.
This can be seen as an arbitrariness in the choice of coordinates in $\fields$.
There is a similar arbitrariness in the choice of coordinates
on $\theories$, due to the freedom of redefining the couplings $g_i$.
We will restrict the group $\redef$ of allowed field redefinitions
in such a way that the transformed action belongs to the
same class of functionals as the original action.
Since $\Gamma_k$ is the most general functional in this class,
given any field redefinition $\phi'=\phi'(\phi)$,
one can find a new set of couplings $g'_i$ such that, for fixed $k$,
$$
\Gamma_k(\phi'(\phi),g)=\Gamma_k(\phi,g')\ .\eqno(4)
$$
This defines an action of $\redef$ on $\theories$. 
We are free to choose a coordinate system on $\theories$
which is adapted to these trasformations, in the sense that
a subset $\bar g_{\bar \imath}$ of couplings are invariant under the action
of the group while a subset $\hat g_{\hat \imath}$ transform nontrivially.
The couplings $\hat g_{\hat \imath}$ are called redundant or inessential,
while the couplings $\bar g_{\bar \imath}$ are called essential.
(It is important to stress that this distinction is parametrization--dependent;
in Appendix A I illustrate this point with a few specific examples.)
A theory will generally contain infinitely many essential and infinitely many
inessential parameters, and all physical observables can be expressed as functions
of the essential couplings only.

One can exploit reparametrization invariance
to eliminate the redundant couplings and follow only the
flow of the essential ones. In general in an effective QFT
there will still be infinitely many couplings to follow, 
but in approximation schemes where only a finite number of couplings
is retained this may be a useful way of reducing the number
of variables.
Field redefinitions also play an important role in discussions of composite
operators, see e.g. [8-10]. 
Mathematically, the ``basic''  RG flow defined on the space $\theories$
projects onto a flow in the quotient space $\bar\theories=\theories/\redef$.
In practice, the flow on $\bar\theories$ can be described explicitly
by giving the complete flow of the essential couplings.
It is characterized by beta functions
which differ from the ``basic'' beta functions by suitable infinitesimal
field redefinitions, as discussed in the introduction.
Such redefinitions are proportional to the beta functions
of the redundant couplings and are usually written in terms
of the anomalous dimensions, defined as
$$
\eta_{\hat \imath}={\beta_{\hat \imath}\over g_{\hat \imath}}
=\partial_t\log \hat g_{\hat \imath}\ .\eqno(5)
$$

Of all field redefinitions, the ones that will interest us in this paper are 
the constant rescalings.
It is generally the case that for each field $\phi_A$ 
(or more precisely for each multiplet of fields) 
there is a scaling invariance of the action, 
provided also the couplings are suitably transformed.
By means of this scaling, for each field multiplet 
one can eliminate one coupling from the action. 
The usual choice is to eliminate the wave function renormalization 
constant $Z_A$,
which is the coefficient of the kinetic term of the field multiplet.

In Quantum Gravity, the RG flow has some special features that are
not present in other QFTs.
This is due to the special meaning of the metric, which is the field used
to measure lengths.
As in any theory, the action is invariant under rescalings of the
metric, provided all fields and couplings are also rescaled suitably:
$$
S(g_{\mu\nu},\phi_A;g_i)=
S(b^{-2}g_{\mu\nu},b^{d_A}\phi_A;b^{d_i}g_i)\ .\eqno(6)
$$
In this equation, $d$ is the canonical dimension of each field or coupling, 
in units of mass.
Thus, dimensional analysis is a statement of invariance of the action under 
rescalings of the metric.
Note that this definition of canonical dimension holds whether or not the metric 
is treated as a dynamical variable, but the freedom to redefine the metric really
only arises in quantum gravity.
This $b$-invariance can be seen as the mathematical manifestation of the fact 
that dimensionful quantities cannot, strictly speaking, be measured: 
one can only measure dimensionless ratios of dimensionful quantities. 
By a measurements of a dimensionful quantity one really means
the measurement of the ratio of the dimensionful quantity and a suitably defined unit.
To choose a system of units is to break the scale invariance of dimensional analysis.

Scale invariance is usually assumed to be broken in QFT, due to the need to introduce
the mass scale $k$ (or in other approaches, an UV cutoff or a renormalization point). 
However, if $k$ is rescaled by a factor $b$
(as its mass dimension requires), scale invariance is maintained also in the quantum theory:
$$
\Gamma_k(g_{\mu\nu},\phi_A;g_i)=
\Gamma_{bk}(b^{-2}g_{\mu\nu},b^{d_A}\phi_A;b^{d_i}g_i)\ .\eqno(7)
$$
If one did not transform the cutoff $k$, then there could be an anomaly.
The two situations correspond therefore to two different realizations of the same
abstract group of scale transformations. 
Dimensional analysis is based on the realization where scale invariance is preserved.

As mentioned in the introduction, the standard definition of the RG
in QFT involves (ii) a constant rescaling of the field and (iii) a constant
rescaling of coordinates and momenta.
A rescaling of space can be treated equivalently as a rescaling of
the metric; in the former case, one is implicitly assuming that the coordinates
have dimension of length and $\gmn$ is dimensionless, while in the
latter the coordinates are dimensionless, as is more natural
in General Relativity, and the metric has dimension of length squared.
There follows that in the presence of dynamical gravity step (iii)
can be seen naturally as a special case of step (ii) applied to
the metric field.
Furthermore, in the case of pure gravity, the rescalings (ii) and (iii)
are not independent operations, and as a result one can use them to
eliminate only one coupling, instead of two, as is the case in ordinary
QFTs [11].

A related point is that in the case of gravity, the group $\redef$
of field redefinitions affects also the parameter $k$.
It is therefore convenient to treat $k$ as one of the couplings of the theory.
Formally, we can think that the space $\theories$ contains a factor
$R^+$ parametrized by $k$.
When $k$ is treated as a coupling, we can ask whether it is essential.
As discussed above, this will depend on the choice of parametrization of $\theories$.
Due to the scaling invariance (7) there will always be at least one redundant 
(dimensionful) coupling, which can be eliminated by a choice of units.
The standard choice in QFT is to take $k$ as unit of mass.
This means choosing coordinates $\{k,\tilde g_i\}$, where 
$\tilde g_i=g_i k^{-d_i}$ are the couplings measured in units of the cutoff.
In this parametrization, $k$ is a redundant coupling and the $\tilde g_i$
are invariant under the rescalings (7).
With this choice $k$ never appears explicitly in any physical observable,
nor in the beta functions, so that the RG flow is given by an autonomous
system of differential equations.

Alternatively, we can also pick any coupling $h$ of dimension $d\not=0$
and take $h^{1/d}$ as unit of mass.
Then we choose coordinates $\{h,\hat k,\hat g_i\}$, 
where $\hat k=k h^{-(1/d)}$ and $\hat g_i=g_i h^{-(d_i/d)}$
are the cutoff and the other couplings measured in units of $h$.
In this parametrization $h$ is redundant and the other, dimensionless
coordinates are invariant under the rescalings (7). Which ones of them
is essential then depends upon the way in which the remaining
field reparametrizations act; however, in these units $\hat k$ is essential
and physical observables may depend on it.

\medskip
\leftline{\bf 3. RG and Systems of Units in a toy model}
\smallskip
In this section I will illustrate various forms of the RG
in a toy model consisting of gravity coupled to a real scalar ``Higgs'' field $\phi$.
Besides serving as an illustration of the general concepts described
in the previous section, such a model also contains
the main ingredients that enter into the hierarchy problem.
While very incomplete in some respects, it may still be
sufficiently realistic to capture some of the physical effects
that play a role in the generation of the hierarchy.
It is also useful for the comparison with [5].

We will deal with actions of the general form (metric signature is $-+++$)
$$
S(\gmn,\phi;\Lambda,Z_g,Z_\phi,m_H,\lambda,\xi)=
\int d^4x \sqrt{g}\left[\Lambda+Z_g R[g]
-{Z_\phi\over 2}g^{\mu\nu}\partial_\mu\phi\partial_\nu\phi
-{1\over 2}\lambda (\phi^2-\upsilon^2)^2
+\xi\phi^2 R[g]
\right]\ .\eqno(8)
$$
Here $Z_g={1\over 16\pi G}$ is, in the linearized theory, the wave function renormalization of the graviton.
The potential $V$ is assumed to have the familiar quartic form with nonvanishing minimum.
It can be parametrized by the value of the VEV, $\upsilon$,
by the Higgs mass $m_H^2={d^2V\over d\phi^2}|_{\phi=\upsilon}$
and the quartic coupling
$\lambda={1\over 12}{d^4V\over d\phi^4}|_{\phi=\upsilon}$.
These parameters are related by $m_H^2=4\lambda\upsilon^2$,
so we can choose $m_H$ and $\lambda$ as independent parameters
and treat $\upsilon$ as a function of the other two.

Our arguments in the following sections will be based on the assumption that 
in the relevant range of energies the effective action $\Gamma_k$ can be
well approximated by a functional of the form (8),
where the couplings $\Lambda$, $Z_g$, $Z_\phi$, $m_H^2$, $\lambda$
and $\xi$ are now $k$-dependent (we omit to write this dependence
explicitly for notational simplicity).
The effective action has the following two scaling properties:
$$
\Gamma_k(\gmn,\phi;\Lambda,Z_g,Z_\phi,m_H^2,\lambda,\xi,\ldots)=
\Gamma_{bk}(b^{-2}\gmn,bc^{-1}\phi;
b^4\Lambda,b^2 Z_g,c^2Z_\phi,b^2 c^2 m_H^2, c^4\lambda,c^2\xi,\ldots)
\eqno(9)
$$
where $b$ and $c$ are positive real numbers and
the dots stand for higher order terms that we are neglecting.
The transformations with parameter $c$ express the behaviour of
the theory under rescalings of the scalar field.
The transformations with parameter $b$ are the expression of dimensional analysis
and are a special case of (7).

Because the action (8) has two scaling invariances, one can use them eliminate
two combinations of the couplings from the action.
The parameter $c$ can be used to fix any coupling that multiplies
an operator containing the field $\phi$. The standard choice is to fix $Z_\phi=1$.
We will always stick to this choice in the following (for an alternative see Appendix A).
The parameter $b$ can be used to fix any one of the dimensionful couplings.
By definition, that coupling then defines a unit of length and mass.
Let us now consider various choices of units.
\smallskip
\goodbreak
\leftline{\it Cutoff units}
In RG theory the standard choice is to take the cutoff $k$ as a unit of mass. 
By means of a scaling (9) with parameters $b=1/k$, $c=1/\sqrt{Z_\phi}$
one can define the normalized action in cutoff units
$$
\eqalign{
\tilde\Gamma(\tilde g_{\mu\nu},\tilde\phi;\tilde\Lambda,\tilde Z_g,\tilde 
m_H^2,\tilde\lambda,\tilde\xi,\ldots)
&=\Gamma_1(\tilde g_{\mu\nu},\tilde\phi;\tilde\Lambda,\tilde Z_g,1,\tilde 
m_H^2,\tilde\lambda,\tilde\xi,\ldots)\cr
&=\Gamma_k(g_{\mu\nu},\phi;\Lambda,Z_g,Z_\phi,m_H^2,\lambda,\xi,\ldots)\ ,}\eqno(10)
$$
where
$$
\tilde g_{\mu\nu}=k^2 g_{\mu\nu}\ ;\ \
\tilde\phi={\sqrt{Z_\phi}\over k}\phi\ ;\ \
\tilde \Lambda={1\over k^4}\Lambda\ ;\ \ 
\tilde Z_g={1\over k^2}Z_g\ ;\ \ 
\tilde m_H^2={1\over Z_\phi k^2}m_H^2\ ;
\tilde \lambda={1\over Z_\phi^2}\lambda\ ;\ \
\tilde \xi={1\over Z_\phi}\xi\ ,\ldots
\eqno(11)
$$
In this parametrization $k$ and $Z_\phi$ are inessential and do not appear
among the arguments of $\tilde\Gamma$; the anomalous dimension
of the scalar field is defined according to (5) as 
$\eta_\phi \equiv \eta_{Z_\phi}=\partial_t\log Z_\phi$.
As noted in [11], there is not enough freedom to eliminate
$k$ and $Z_g$ at the same time, so with this choice of units
the action $\tilde\Gamma$ (and also the beta functions, and physically measurable quantities)
depends explicitly on $\tilde Z_g={1\over 16\pi\tilde G}$,
where $\tilde G=k^2 G$ can be viewed as Planck's area measured in cutoff units.

An infinitesimal RG transformation of $\tilde\Gamma$ then consists of: 
\item{1)} functional integration over a shell of
momenta, from $k$ to $k+\delta k$, with $\delta k<0$.
This produces a variation of the couplings $\delta g_i=\beta_i\delta t$.
Since $k$ has changed by a factor $1+\delta t$ and
$Z_\phi$ has changed by a factor $1+\eta_\phi\delta t$,
this has to be followed by
\item{2)} a rescaling as in (9) with parameter $b=1-\delta t$ and 
\item{3)} a rescaling as in (9) with parameter $c=1-{1\over2}\eta_\phi\delta t$, 
restoring the value $Z_\phi=1$.

\smallskip
\leftline{\it Planck units}
Another possibility is to choose $\sqrt{16\pi Z_g}=G^{-1/2}=m_P$ as the unit of mass.
Being based on the fundamental constants that enter in special relativity,
in the quantum theory and in the theory of gravity, these units are often
considered ``the most fundamental ones'' [2].
A transformation with parameters $b=1/\sqrt{16\pi Z_g}=\sqrt{G}$, $c=1/\sqrt{Z_\phi}$
can be used to define the action in Planck units:
$$
\eqalign{
\Gamma'_{k'}(g'_{\mu\nu},\phi';\Lambda',m_H^{\prime 2},\lambda',\xi',\ldots)
&=\Gamma_{k'}(g'_{\mu\nu},\phi';\Lambda',{1\over 16\pi},1,m_H^{\prime 2},\lambda',\xi',\ldots)\cr
&=\Gamma_k(g_{\mu\nu},\phi,\Lambda,Z_g,Z_\phi,m_H^2,\lambda,\xi,\ldots)\ ,}\eqno(12)
$$
with
$$
k'={k\over\sqrt{16\pi Z_g}};\ 
g'_{\mu\nu}=16\pi Z_g\gmn;\ 
\phi'=\sqrt{Z_\phi\over 16\pi Z_g}\phi;\ 
\Lambda'={\Lambda\over (16\pi Z_g)^2};\ 
m_H^{\prime 2}={m_H^2\over 16\pi Z_g Z_\phi};\ 
\lambda'={\lambda\over Z_\phi^2};\ 
\xi'={\xi\over Z_\phi}\ .\eqno(13)
$$
As noted in [11], there is not enough freedom to eliminate
$k$ and $Z_g$ at the same time, so with this choice of units
the action $\Gamma'$ depends explicitly on $k'$,
which can be viewed as the cutoff measured in Planck units.
This is rather unusual, because the flow equations become
non--autonomous.
Let us denote $\eta_g \equiv \eta_{Z_g}=\partial_t\log Z_g$ the 
anomalous dimension of the graviton.

An infinitesimal RG transformation of $\Gamma'$ consists of: 
\item{1)} functional integration over a shell of momenta, as above;
since the couplings $Z_g$ and $Z_\phi$ are modified, this is followed by 
\item{2)} a rescaling as in (9) with parameter $b=1-{1\over 2}\eta_g\delta t$, and
\item{3)} a rescaling as in (9) with parameter $c=1-{1\over 2}\eta_\phi\delta t$, 
restoring the unit value $Z_\phi$.

\smallskip
\leftline{\it Higgs units}
A third possibility is to use the Higgs mass as unit of mass;
we will call these ``Higgs units''.
Note that while the Higgs mass is not known,
the mass of the electron, which differs from the Higgs mass
by the ratio of the coupling $\lambda$ to a Yukawa coupling,
is at the basis of atomic spectroscopy and hence of modern metrology.
Thus, the Higgs units defined here are conceptually closely related to ordinary metric units.
By means of a transformation as in (9), with parameters 
$b=\sqrt{Z_\phi}/m_H$ and $c=1/\sqrt{Z_\phi}$ 
we can define the action in Higgs units
$$
\eqalign{
\Gamma''_{k''}(\gmn'',\phi'';\Lambda'',Z''_g,\lambda'',\xi'',\ldots)
&=\Gamma_{k''}(\gmn'',\phi'';\Lambda'',Z''_g,1,1,\lambda'',\xi'',\ldots)\cr
&=\Gamma_k(\phi,g_{\mu\nu},\Lambda,Z_g,Z_\phi, m_H^2,\lambda,\xi,\ldots)\ ,}\eqno(14)
$$
with 
$$
k''={\sqrt{Z_\phi} \over m_H}k\ ;\ \
g''_{\mu\nu}={m_H^2\over Z_\phi}\gmn\ ;\ \
\phi''={Z_\phi\over m_H}\phi\ ;\ \
\Lambda''={Z_\phi^2\over m_H^4} \Lambda\ ;\ \
Z''_g= {Z_\phi \over m_H^2}Z_g\ ;\ \
\lambda''={1\over Z_\phi^2}\lambda\ ;\ \ 
\xi''={1\over Z_\phi}\xi\ ,\ldots\ \
\eqno(15)
$$
Again, the action depends explicitly on $k''$, which is the cutoff measured
in Higgs units.

An infinitesimal RG transformation of $\Gamma''$ consists of: 
\item{1)} functional integration over an infinitesimal shell of momenta, as above.
Since this modifies both $m_H^2$ and $Z_\phi$, it is followed by
\item{2)} a rescaling as in (9) with parameter 
$b=1-{1\over 2}\eta_{m_H^2}\delta t+{1\over 2}\eta_\phi\delta t$, 
where $\eta_{m_H^2}$ is defined as in (5), and 
\item{3)} a rescaling as in (9) with parameter $c=1-{1\over 2}\eta_\phi\delta t$, 
restoring the unit value $Z_\phi$.
\smallskip

This list does not exhaust all possibilities.
For example, one could use the cosmological constant as unit of mass.
This is not a practical choice, because the cosmological constant
is not known to any good approximation, but it is a rather
natural choice in principle (see eqs, (A.5-9) in the Appendix)
and it is used in lattice calculations (see \eg\ [12]).
I leave it to the reader to work out the transformations of the
fields and couplings in this case.

\smallskip
It is important to underline that the use of these complete RG transformations
on $\bar\theories$ 
is entirely equivalent to working with the ``basic'' RG transformations on $\theories$.
For example the complete RG transformation of the scalar mass parameter $m_H^2$
in cutoff units is
$$
\tilde\delta m_H^2=\beta_{m_H^2}\delta t-2 m_H^2\delta t-\eta_\phi m_H^2\delta t\ .\eqno(16)
$$
The extra terms which come from steps 2 and 3 cancel out when we compute the 
transformations of the dimensionless couplings listed in eq. (11).
For example, in the case of the mass, using the definition of $\tilde m_H^2$ in (11),
since $\tilde\delta k=0$ and $\tilde\delta Z_\phi=0$ by definition, we find
$$
\delta\tilde m_H^2={\tilde\delta m_H^2\over Z_\phi k^2}=\tilde\delta\tilde m_H^2\ ,\eqno(17)
$$
which means that the complete RG transformation of $\tilde m_H^2$
is given just by the ``basic'' RG transformation, provided it is applied also
to the factors $k^2$ and $Z_\phi$ appearing in the denominator.

Similar remarks apply to the case of the other systems of units.
For example in Planck units the complete transformation of the mass is
$$
\delta' m_H^2=\beta_{m_H^2}\delta t-\eta_g m_H^2\delta t-\eta_\phi m_H^2\delta t\ .\eqno(18)
$$
In Planck units the total transformations $\delta'Z_g=0$ and $\delta'Z_\phi=0$ by design, 
so the complete transformation of the dimensionless ``mass in Planck units'' 
$m_H^{\prime 2}$ defined in (13) is
$$
\delta m_H^{\prime 2}={\delta' m_H^2\over 16\pi Z_g Z_\phi}=\delta' m_H^{\prime 2}\ .\eqno(19)
$$
meaning that the complete RG transformation of $m_H^{\prime 2}$ 
can be computed just from
the ``basic'' RG transformation, provided it is applied also to the factors
$Z_g$ and $Z_\phi$ in the denominator.
Similar considerations apply also to the other primed and double primed couplings.

The gravitational RG has been studied intensively in the last ten years with the
main aim of proving nonperturbative renormalizability of the theory
along the lines of [13].
The derivations of the beta functions for gravity in the Einstein--Hilbert
truncation [14] and gravity coupled to matter in [15], 
as well as their applications in [16,17] to the search of a gravitational FP, 
were based on the exact RG equation, taking into account only
the first of the three steps described above for each choice of unit.
The same applies to subsequent calculations that take into account
higher powers of the curvature [18].
By the arguments given above, the beta functions found in these papers
for the dimensionless ratios $\tilde G=Gk^2$ and $\tilde\Lambda=\Lambda/k^4$
are the complete beta functions in cutoff units and
do not need any further correction term (in particular, the FP is unaffected).

Let us conclude this section by observing that the primed (Planck) and
double primed (Higgs) units could be ill-suited for the discussion of FP's. 
The great virtue of cutoff units is that the vectorfield defined by the
beta functions is $t$--independent.
This is not the case with the other choices of units, since, as noted above, 
in these units the beta functions have an
explicit dependence on the cutoff variables $k'$ and $k''$ respectively.
This makes the physical interpretation of the system of equations
less transparent [11].

\medskip
\leftline{\bf 4. The beta functions}
\smallskip
Having discussed the general form of the RG equations of the toy model 
in various systems of units, 
I will now illustrate the actual behaviour of the couplings,
under the assumption that the action (8) gives a good description of physics
at the energy scale that is of interest.
Beyond this fundamental assumption,
any calculation will require a number of approximations.
First, we are going to neglect the anomalous dimension $\eta_\phi$.
This is supported by the calculations of [19], where 
it was shown that the anomalous dimension is suppressed by
some power of $k/m_P$ below the Planck scale.
Furthermore, in the UV limit we see from eqn.(6.1a) of [19],  
together with the results of [16,17],
that $\eta_g$ tends to a constant of order 10$^{-3}$ if there is a FP.
From here on we shall simply assume $Z_\phi=1$ and $\eta_\phi=0$.
In accordance with observations, we will also assume that the 
cosmological constant is much smaller than all other scales that come into play.
Since our main focus is not on explaining the smallness of the cosmological constant, 
we shall simply assume that this condition is met.

Qualitatively, the overall picture is characterized by the presence of
two thresholds which correspond to the physical Higgs mass and the Planck mass.
There are therefore three distinct regimes: 
the low energy regime $\sqrt{\Lambda}\ll k^2 \ll m_H^2 \ll Z_g$, 
the intermediate regime $\sqrt{\Lambda}\ll m_H^2 \ll k^2 \ll Z_g$,
and the asymptotic UV regime where, in principle, $k^2\gg Z_g$.
The beta functions of the couplings in (8) have been calculated in [17].
They consist of integrals of certain rational functions of the
momenta, the cutoff and the couplings.
The momentum integrals are peaked at momenta of order $k$ and
in each regime, one can pick in the beta functions
the dominant terms and discard all the others.
This leads to dramatic simplifications.

Let us discuss each regime in turn, in order of increasing energy.
In the low energy regime we have
$\beta_{Z_g}=a_1 k^2$, with $a_1=-{29\over 384\pi^2}$,
when using a suitable cutoff function.
The solution of the flow equation has the form
$$
Z_g=\bar Z_g\left(1+{1\over 2}a_1 {k^2\over \bar Z_g}\right)\ ,\eqno(20)
$$
where $\bar Z_g=Z_g(k=0)$.
By definition in this regime the initial value $\bar Z_g$ is much larger than $k^2$
so the second term in (20) is very small with respect to the first
and we can approximate $Z_g$ by the constant $\bar Z_g$.
The anomalous dimension $\eta_g$ is consequently also very small.
When we look at the beta functions of the other couplings,
they are also suppressed by powers of $k^2/Z_g$, so we
can effectively approximate $m_H^2$ by a constant $\bar m_H^2$
and $\lambda$ by a constant $\bar \lambda$.
Altogether, in the low energy regime there is no significant running
of the couplings.

Eventually, as $k^2$ grows, it overcomes the threshold at $k^2\approx \bar m_H^2$.
In the intermediate regime where $\bar m_H^2\ll k^2 \ll Z_g$,
the beta function of $Z_g$ is the same as in the low energy regime,
except for the numerical coefficient $a_1={-33+24\xi\over 384\pi^2}$.
The difference is due to the contribution of 
the scalar field, which is felt only above the mass threshold.
Since $k^2$ is still much smaller than $Z_g$, we can still assume that
$Z_g$ is almost constant and equal to $\bar Z_g$.
In the intermediate regime the dominant term in the beta function of the mass is
$\beta_{m_H^2}=c_1 \lambda k^2$, with $c_1={3\over 4\pi^2}$. 
This is essentially the one-loop result for a quartically self-interacting
scalar theory; the gravitons do not contribute significantly to this beta
function in this regime.
If we assume that the threshold occurs exactly at $k^2=\bar m_H^2$, 
matching the solution in the intermediate regime to the
one in the low energy regime we have
$$
m_H^2(k)=\bar m_H^2\left(1+{1\over 2}\lambda c_1{k^2-\bar m_H^2\over \bar m_H^2}\right)
\approx{1\over2}c_1\lambda k^2\ .\eqno(21)
$$
This running is a reflection of the quadratic
divergences that occur in perturbative calculations.
The dominant term in the beta function of $\lambda$ is
$\beta_\lambda=d_1\lambda^2$ with $d_1={9\over 8\pi^2}$.
The contribution of gravitons is negligible.
The solution is
$$
\lambda(k)={\bar\lambda\over 1-d_1\bar\lambda\ln{k\over\bar m_H}}\ ,\eqno(22)
$$
where $\bar\lambda=\lambda(k=\bar m)$ is the value of $\lambda$
at (and below) the threshold.
This is the standard one loop result, exhibiting a Landau pole
at $k=\bar m_H e^{1/(d_1\bar\lambda)}$. 
The pole can be avoided if 
$d_1\bar\lambda < 1/\ln(Z_g/\bar m_H)$.

Beyond the Planck scale, field theory will make sense
if the couplings approach a FP [13].
In this case, it never really happens that $k^2\gg Z_g$; instead, $Z_g$ scales exactly like $k^2$,
so that the ratio $\tilde Z_g=Z_g(k)/k^2$ tends to a constant.
In [17] it was found that the model (8) has a ``Gaussian--matter FP'' 
with nonzero values $\tilde\Lambda_*$ and $\tilde G_*$ while all other couplings are zero.
It is useful to emphasize at this point, however, that the subsequent discussion
does not depend on the existence of such a FP.
The RG behaviour we have encountered in the intermediate regime would still remain
valid even if the field theoretic description was only an effective one,
with a physical cutoff somewhere near the Planck scale.

Let us now summarize the consequences of these findings for the parameter $\conv$.
In the low energy regime there is no significant running of any quantity,
so Higgs units and Planck units simply differ by a fixed multiplicative factor
$$
\bar\conv={\bar m_H\over \sqrt{16\pi\bar Z_g}}=\bar m_H\sqrt{\bar G}\approx 10^{-16}\ .\eqno(23)
$$
This is the regime where all of current particle physics takes place.
In the intermediate regime, $Z_g$ is still essentially constant but $m_H^2$
runs, to leading order like $k^2$.
From eq.(21), keeping only the leading terms, we find
$$
\conv\approx\sqrt{c_1\lambda\bar G\over2} k\ .\eqno(24)
$$

In the asymptotic UV regime, assuming that the theory approaches
an UV FP, $\conv$ would tend again to a constant
$$
\conv_*={\tilde m_{H*}\over \sqrt{16\pi\tilde Z_{g*}}}
=\tilde m_{H*}\sqrt{\tilde G_*}\ ,\eqno(25)
$$
where the asterisk denotes the values of the variables at the FP.
The value of $\conv_*$ clearly depends on the details of the FP.
In [17] we searched for a FP with nonzero values of the scalar couplings,
but we were unable to find one.
It seems likely that if a potential exists with nonzero mass, it is
not polynomial. However, this issue remains unsettled.
In the case of the ``Gaussian--Matter FP'', after reaching a maximum around the Planck
scale, the ratio $\conv$ tends to zero for $t\to\infty$.

To conclude this section, some cautionary comments are in order.
It is not granted that the action (8) is a good description
of the world at any energy scale, much less that is a good description
at {\it all} energy scales.
In fact, experience with other QFTs leads us to expect that if
(8) has anything to do with the real world in the UV regime, then
it is unlikely to do so at lower energies.
In this respect QCD is probably a useful guide: it has a very simple
description in terms of quark and gluon fields at high energy, near its
UV FP, but this description becomes practically useless at energies of the
order of 1 GeV, where the strong coupling becomes strong and other terms
in the action become important.
Likewise, one would expect that if a gravitational FP is governed 
by an action like (8), perhaps containing a few other terms,
then at and below the Planck scale the action, in terms of the
variables $g_{\mu\nu}$ and $\phi$, will be an extremely complicated,
possibly nonlocal functional.

This is in contrast to the common expectation that a functional like (8)
may be a reasonable approximation in the low energy and possibly
in the intermediate regime
(this expectation comes from the fact that the gravitational interaction
is proportional to the external momenta in Feynman diagrams, 
so that perturbation theory works well with the action (8) at
sub--Planckian energies).
It is possible that this expectation is correct and that there is no UV FP.
Then, the preceding analysis could still be applied in the 
low energy and intermediate regime.
Alternatively, a possible reconciliation of these points of view may come through a
field redefinition: at intermediate/low energies the action may again have
a general structure similar to (8),
but in terms of other variables $\gamma_{\mu\nu}$ and $\varphi$,
related to $g_{\mu\nu}$ and $\phi$ by some functional field redefinition
(note that one always expects to have some metric in the action).
Examples of such transformations are already well--known in scalar--tensor 
theories [20]. This would parallel the transition from quark/gluon to
meson/baryon degrees of freedom in QCD at low energy [9].

\medskip
\leftline{\bf 5. AdS space, from the bottom up}
\smallskip
In the preceding section we have seen the ``basic'' beta functions on $\theories$
for the toy model.
Let us now see what the complete RG flow on $\bar\theories$ looks like
for various choices of units.
This will allow us to make contact with the RS model.

For simplicity we concentrate on the second, third and fourth terms of the action (8),
which carry all the relevant information. 
To simplify the discussion, we also assume that the three regimes described in section 4
extend all the way to the thresholds.
We choose $t=0$ at the Planck scale, so that $t=\log(k/m_P)$.
Let us assume that at the Planck scale
the scalar mass has some ``natural'' value of order of the Planck mass:
$m_H^2(t=0)\approx Z_g$.
In the rest of this section we will consider only the leading scaling behaviour
of each quantity in the intermediate regime, 
neglecting the anomalous dimensions $\eta_g$ and $\eta_\phi$,
and the logarithmic running of $\lambda$.

In cutoff units at the Planck scale the action reads
$$
\tilde\Gamma\Big|_{t=0}=
\int d^4x \sqrt{\tilde g}\left[\tilde Z_g R[\tilde g]
-{1\over 2}\tilde g^{\mu\nu}\partial_\mu\tilde\phi\partial_\nu\tilde\phi
-{1\over 2}\tilde\lambda(\tilde\phi^2-\tilde\upsilon^2)^2
\right]
\eqno(26)
$$
All couplings are evaluated at $t=0$. Let us now run the RG in cutoff units,
as described in Section 3, towards lower energies (negative $t$).
As discussed in Section 4, $Z_g$ does not run significantly; 
for the mass we keep only the term proportional to $k^2$
and neglect any constant. Then we have
$$
\tilde Z_g(t)={\bar Z_g\over k^2}=\tilde Z_g(0)e^{-2t}\ ;
\ \ \tilde m_H(t)^2=\tilde m_H(0)^2\ .\eqno(27)
$$
Since we neglect the running of $\lambda$, also $\tilde\upsilon(t)^2=\tilde\upsilon(0)^2$.
Furthermore, since $\eta_\phi\approx 0$, the third step of the
RG transformation does not change the fields.
However, the second step of the RG transformations is nontrivial: it 
generates a rescaling of the metric and of the scalar field with a suitable power
of the factor $1-\delta t$ at each RG step:
$$
{d \tilde g_{\mu\nu}\over dt}=2 \tilde g_{\mu\nu}\ ;\quad\quad
{d \tilde\phi\over dt}=- \tilde\phi \ ,\eqno(28)
$$
which yields
$$
\tilde g_{\mu\nu}(t)=e^{2t}\tilde g_{\mu\nu}(0)\ ;\qquad\qquad
\tilde\phi(t)=e^{-t}\tilde\phi(0)\ ,\eqno(29)
$$
This rescaling has to be applied to all the unintegrated Fourier modes of the
field, having momenta lower than $k$. 
In practice we are mostly interested in the zero momentum modes
(the constant scalar and a flat metric)
and (29) implies that these modes have to be rescaled with $t$ as shown.

Taking into account equations (27,29), the action at some lower scale $t<0$ is given by 
$$ 
\eqalignno{ \tilde\Gamma_{t}=& \int d^4x
\sqrt{\tilde g(t)}\left[\tilde Z_g(t) R[\tilde g(t)] 
-{1\over 2}\tilde g^{\mu\nu}(t)\partial_\mu\tilde\phi(t)\partial_\nu\tilde\phi(t)
-{1\over 2}\tilde\lambda(t)(\tilde\phi(t)^2-\tilde\upsilon(t)^2)^2 \right]\cr 
=&\int d^4x \sqrt{\tilde g(0)}\left[\tilde Z_g(0)R[\tilde g(0)] 
-{1\over 2}\tilde g^{\mu\nu}(0)\partial_\mu\tilde\phi(0)\partial_\nu\tilde\phi(0) 
-{1\over2}\tilde\lambda(0)(\tilde\phi(0)^2-e^{2t}\tilde\upsilon(0)^2)^2 \right]\ .&(30)\cr 
} 
$$

In Planck units the details of the calculation differ but the final result is the same. 
The action at the Planck scale is given by 
$$
\Gamma'\Big|_{t=0}= 
\int d^4x \sqrt{g'}\left[{1\over16\pi}R[g'] 
-{1\over 2}g'^{\mu\nu}\partial_\mu\phi'\partial_\nu\phi' -{1\over2}\lambda'(\phi'^2-\upsilon'^2)^2 \right]\ . \eqno(31) 
$$ 
In these units, by definition 
$Z'_g={1\over 16\pi}$ independent of $t$.  Since we neglect $\eta_g$,
the second step of the RG is trivial.  
Since we neglect $\eta_\phi$, the third step of the RG is trivial.  
The mass runs like 
$$
m'_H(t)^2=m'_H(0)^2 e^{2t}\ .\eqno(32) 
$$ 
Everything else, including field normalizations, is $t$-independent within the approximations
that we made, so in these units the scale-dependence of the mass is immediately evident.
The reader can check that, due to linear running of the mass, 
Higgs units work like cutoff units.

We can then compare these results to the original RS model.
One sees that the calculations leading from eq.(26) to eq.(30)
are identical to the calculations leading from eq.(17) to eq.(19) in [5].
This should perhaps not come as a surprise, in view of the ``holographic''
interpretation of the RS model [7].
Their five-dimensional metric contains a one-parameter family of four-dimensional
metrics with an exponentially varying warp factor:
$$
\gmn(x,t)=e^{at}\eta_{\mu\nu}\ .\eqno(33)
$$
We see that essentially the same one-parameter family of metrics 
is generated in a purely four-dimensional theory between the Higgs and the Planck scales,
when one works in cutoff units (or in Higgs units).

To make the connection more complete, we should say how
distances have to be defined in the space $R^+$ parametrized by the scale $k$.
Since $k$ is typically identified with some momentum variable appearing in
a process, it may seem natural to choose the ``proper distance''
in cutoff space to be proportional to $dk$, as would befit a linear variable.
But the space $R^+$ is not a linear space and this is not the natural choice.
To motivate this from a physical point of view, observe that
the units that we have chosen are subject to RG flow, so the question arises: 
when we measure a mass $m$ in a certain system of units, 
at what scale should the unit be taken?
The natural, or ``intrinsic'', definition is to take $k=m$.
This is the only way to define a real number out of a mass and a running unit,
without making reference to another mass scale.
Thus, when we ``measure'' a physical mass $m$
in Higgs units the real number that is obtained is the ratio
$$
{m\over m_H(m)}\ .\eqno(34)
$$
The physical rationale of this assumption is that when we perform a measurement 
of a mass parameter, we are not really measuring just that mass,
but also the unit of mass. The two measurements cannot be disentangled
and all one obtains is a value for the ratio between the two;
if $m$ is the only scale in the system, (34) follows.
Similarly a measurement of a length $\ell$ gives the number ${\ell m_H(\ell^{-1})}$.
On the other hand a measurement of a mass $m$ in Planck units gives the number
$$
{m \sqrt{G(m)}}\ ,\eqno(35)
$$
where Newton's constant is evaluated at the scale $k=m$,
while a measurement of a length $\ell$ yields
$\ell/\sqrt{G(\ell^{-1})}$,
where $G$ is evaluated at the scale $k=\ell^{-1}$.

This natural assumption leads to unfamiliar consequences.
Let $\ell_1$ and $\ell_2$ be two lengths and
$r=\ell_2/\ell_1$ their dimensionless ratio.
The ratio of the two real numbers which give the values of
the lengths in the two systems of units 
will not in general be equal to each other, nor to $r$.
According to our postulate, the ratio of the two lengths 
in Planck units is:
$$
r_P={\ell_2\over\sqrt{G(\ell_2^{-1})}}
{\sqrt{G(\ell_1^{-1})}\over \ell_1}=
\sqrt{G(\ell_1^{-1})\over G(\ell_2^{-1})}\ r\ .\eqno(36)
$$
Similarly, the ratio of the two lengths in Higgs units, is
$$
r_H={\ell_2 m_H(\ell_2^{-1}) \over \ell_1 m_H(\ell_1^{-1})}=
{m_H(\ell_2^{-1}) \over m_H(\ell_1^{-1})}\ r=
{\conv(\ell_2^{-1}) \over \conv(\ell_1^{-1})}\ r_P
\ .\eqno(37)
$$
So, if the conversion factor $\conv$ is not the same at all scales,
a set of masses that are linearly spaced when referred to Planck units
will not be linearly spaced when referred to Higgs units.
An extreme example of this fact has already been noticed in [11]:
if the QFT of gravity has a fixed point, when $t\to\infty$
the variable $k'$ defined in (13), which gives the value of the cutoff in Planck units,
must tend to a finite limit $k'_*=\sqrt{\tilde G_*}$,
whereas the variable $k''$ defined in (15), which gives the value of the cutoff
in Higgs units, may well tend to infinity.

All this is unusual, but it does not imply any inconsistency.
It only means that cutoff space cannot be thought of as a linear space,
but should be treated as a one-dimensional manifold
\footnote{$^{\rm \S}$}{There may be a connection with so--called
``Doubly Special Relativity'' models, where the Lorentz group is postulated
to act in a nonlinear way on momentum space [21].}
.
The dimensionless variables $t$, $k'$ and $k''$ provide coordinate systems
on this manifold, related to the use of cutoff, Planck and Higgs units respectively.
The relation between these variables is nonlinear and may not even be invertible.
Since by construction $t$ grows as the RG trasformations are iterated,
the direction of growing $t$ defines a natural orientation in cutoff space,
but the directions of growing $k'$ and $k''$ may or may not agree with this orientation
\footnote{${}^{\ddag}$}{For example, it has been observed in [11] that in the approach
to the UV FP, Newton's constant follows damped oscillations.
At each oscillation, there would be two stationary points where $k'$
could not be used as a coordinate, and the direction of growing $k'$
disagrees with the direction of growing $t$ at each second half-cycle.
An alternative coordinate better adapted to the FP behaviour has been proposed in [11].
Similarly in Higgs units, assuming that $\tilde m_H\to 0$ at the FP,
there would be a scale of the order of Planck's scale where $\tilde\beta_{\tilde m_H}$
changes sign. Below this scale $k''$ would be an increasing function
of $t$ and above this scale it would be decreasing.}.

Since cutoff space does not have a natural linear structure, 
a proper notion of distance between, say, the Higgs scale and the Planck scale
requires that a metric be given.
The most natural choice of distance in $R^+$ would be given by counting 
the number of iterations of the RG transformation.
For continuous RG transformations ($\delta t\to 0$) the
proper distance is then simply proportional to $dt$.
Note that this logarithmic scale is also more natural
insofar as it places the infrared limit (infinitely large systems)
at infinite distance from any finite system.

At this point, we have all the ingredients to reconstruct the entire five-dimensional
RS metric from the four-dimensional RG.
Combining the metric on spacetime with the metric on cutoff space, 
together with the assumption that cutoff and physical space are orthogonal,
produces a metric on a five-dimensional manifold.
In the slice between the Higgs and the Planck scales, considering
only the leading behaviour (21), this metric is locally the AdS metric:
$$
ds^2=dt^2+e^{2t}\eta_{\mu\nu}dx^\mu dx^\nu\ .\eqno(38)
$$
In this way one can entirely reconstruct the RS five-dimensional space
from four-dimensional data, without having to postulate that the fifth
dimension is physically accessible.

\medskip
\leftline{\bf 6. Conclusions}
\smallskip
In this paper I have discussed the way in which field rescalings affect
the RG flow in the presence of gravity.
One general lesson that can be drawn from this discussion is that in certain
circumstances the use of dimensionful quantities could be misleading.
The measurement of a dimensionful quantity is in reality the measurement of the ratio
between that quantity and some unit.
In daily life we can refer to a well--established system of units,
which have been chosen for their availability and 
stability under a wide range of circumstances;
in particular, they are unaffected by RG ambiguities.
However, their use in extreme situations and in particular in Quantum Gravity
may require some care.
The relation between a certain dimensionful quantity measured in a high energy
experiment and standard metric units can be very indirect [4].
In order to avoid potential ambiguities, the safest procedure is to consider only dimensionless
ratios, with the understanding that any physical measurement is a simultaneous
measurement of the numerator and the denominator, and that one may not be able to
measure the numerator and denominator separately.
In theoretical calculations, one has to take into account the fact that
both numerator and denominator may be subject to RG flow.

A related point is the following. As long as we study a theory
describing a limited set of physical phenomena, we can always ignore these
issues by taking as a unit something external to the theory. 
For example, if we study only strong interactions we can take the mass of the $Z$ as
an absolute, external unit; if we study only electroweak and strong interactions,
the Planck mass defines an absolute, external unit.
If on the other hand we want to include in the theory also gravitational
phenomena, then also the Planck mass will be subject to RG flow, as we have
seen, and there is no external parameter that can be taken as an absolute unit.
In these circumstances, the safest procedure seems to be to avoid statements
about dimensionful quantities.

To describe physics entirely in terms of dimensionless quantities would be similar to 
describing a gauge theory entirely in terms of gauge invariant quantities,
and could become very cumbersome. However, when dimensionful variables are used,
one should take care to avoid ambiguous statements.
For example, the statement $k^2\ll 1/G$ is a statement about dimensionful quantities
that is unambiguous, because it is equivalent to the statement $k^2G\ll 1$.
On the other hand a statement such as ``$k\to\infty$'' 
(which is usually taken to define the ultraviolet limit) is ambiguous. 
To see why, consider for example the asymptotic UV regime of the toy model, 
which has been briefly mentioned in section 4.
There is evidence for the
existence of a ``Gaussian--matter FP'' where Newton's constant, in cutoff units, 
tends to a nonzero value $\tilde G_*$ and the ratio $m_H/k$ tends to zero [17]. 
This would imply that in the UV limit the cutoff in Higgs units $k''\to\infty$, while
the cutoff in Planck units $k'$ tends to a finite limit $k'_*$.
In such a case saying that ``$k\to\infty$'' is at best misleading.
An unambiguous way of referring to the UV limit could be
to say ``$t\to\infty$'', meaning that the RG transformation
is iterated infinitely many times.
Similarly, to say that a dimensionful coupling such as Newton's constant
is asymptotically free is in itself meaningless.
One has to specify in what units Newton's constant is being measured.
Depending on the flow and on the choice of units,
it could tend to zero, to a finite constant or to infinity
(for example, Newton's constant is always equal to one in Planck units,
so the statement above can evidently never be true in Planck units).

Let us now discuss possible connections with the hierarchy problem.
As mentioned in the introduction, the problem posed by the smallness 
of the ratio (1) in the toy model is conceptually very similar to the 
one posed by the smallness of all lepton and quark masses, in Planck units.
The mass of composite particles poses a different problem.
For example, the mass of the proton is $m_p\approx 10^{-19}m_P$, 
but the origin of this small ratio must be to a large extent 
independent from the hierarchy problem represented by (1).
This is because the mass of the proton originates mostly from the kinetic energy
of the quarks and from the energy of the gluon field that binds them, 
and only in minimal part from the rest mass of its pointlike constituents.
We are now close to having a satisfactory explanation for this problem, 
based on RG effects in QCD [22]. 
The mass of the proton corresponds to the scale 
at which the strong coupling $\alpha_s$ becomes large enough to bind
the quarks. This requires 
$$
\alpha_s(m_p)\approx 1\ .\eqno(39)
$$
Starting from a relatively small value for $\alpha_s$ at the unification scale $m_U$
(which is close to the Planck scale), due to the logarithmic running,
one has to go to extremely small scales before the coupling becomes strong.
To make this a bit more quantitative, solve the RG equation 
$$
{d \alpha_s\over dt}=-b \alpha_s^2\ ,\eqno(40)
$$
where $b={1\over 2\pi}\left(11-{2\over3}n_f\right)$ is the 
beta function in the presence of $n_f$ flavors of quarks, 
and use the condition (39). One gets
$$
m_p=m_U e^{-{1\over b}\left({1\over \alpha_s(m_U)}-1\right)}\ .\eqno(41)
$$
Without entering into details of specific grand unification schemes,
it clear that very small ratios can be generated in this way.

As pointed out before, the masses of hadrons originate from strong interaction
physics and those of pointlike particles from electroweak physics.
The success of the RG of QCD in producing a small ratio for $m_p/m_P$ does not 
imply that a similar mechanism can be found for the masses of the Higgs scalar 
or the electron, but the idea remains extremely attractive.
Insofar as the RS model provides a mechanism for generating small
mass ratios, the ability of the four--dimensional RG to reproduce its
main results seems to be encouraging.
However, to reproduce the RS model from the RG we assumed
that the running mass is proportional to $k$;
more precisely, we have neglected any constant term, which
is permitted in the solution of the RG equation.
When one solves the flow equations starting at $m_U$,
there is no reason to assume that the constant term 
is negligible, so reproducing the trajectory (21)
requires an {\it a priori} fine tuning of the initial
conditions at the Planck or GUT scale.
In the presence of fermion fields there may exist a dynamical mechanism,
based essentially on an approximate IR fixed point for the scalar mass,
that attracts it linearly towards the origin [23].
This mechanism demands a large number of scalar and fermion fields
and/or large values of the Yukawa couplings.
A slow decrease of the Yukawa couplings would then
switch off this running at some small scale, similar to the effect
described above for QCD.
It thus seems worth looking in more detail at the RG
flow of the model, perhaps taking into account other parameters
or other matter fields,
and searching for RG trajectories that exhibit an extended
intermediate regime.
It would be natural to explore these issues under the
assumption of the existence of an UV FP, though this hypothesis
may not be necessary for an RG mechanism to work.

Another extension of this work would be to consider the effect
of more general classes of field redefinitions, for example
those involving conformal rescalings of the metric.
Presumably, one could relate this to extensions of the RS model
where the branes are not parallel [24]. We note that in this
generalization the scale transformations of dimensional analysis
would have the status of gauge transformations.

It is also worth comparing these RG arguments to Dirac's original prediction that $G$,
when measured in atomic units, would decrease like the inverse of the
epoch $\tau$.
One can reasonably assume that in a cosmological context the cutoff $k$
can be identified with the inverse of the epoch [25] (or maybe some function thereof).
Now the value of $G$ in Higgs units is $Gm_H^2=\conv^2$.
Therefore, in the intermediate regime it should behave like $\tau^{-2}$,
and furthermore it should stop running after an epoch of the order $m_H^{-1}$.
So, if this picture is correct, Newton's constant measured in Higgs units
would indeed be time--dependent, but only in the very early universe, 
before the electroweak phase transition.

We have seen that within certain approximations the AdS geometry
can be interpreted as a description of the quadratic divergences
occurring in the toy model.
For a review of holography and its relation to other important current 
developments in theoretical physics see [26].
I would like to emphasize once again that this reconstruction of the AdS metric 
is purely formal, whereas in the RS model it is assumed that it is in principle possible 
(at least for gravitons) to propagate into the fifth dimension.
Furthermore, while RS derive the RG flow from the five-dimensional Einstein equations,
we used a standard four-dimensional beta function calculation.
The phenomenology of the RS model, as well
as that of other higher-dimensional theories that place the Planck scale
close to the Higgs scale, is very different
from the one of the purely four-dimensional model discussed here.
Since $t$ is not interpreted as a physical dimension,
there are no Kaluza-Klein modes and no radion field.
Insofar as these higher-dimensional theories may be amenable to experimental
verification, the four-dimensional RG approach discussed here provides
an experimentally distinguishable alternative.

In conclusion let us observe that until now particle physics
has been explored in the low energy regime, where units
can be assumed not to run significantly.
Particle accelerators are now on the verge of entering
into the intermediate regime, where the metric in Higgs units
($g''_{\mu\nu}$) will begin to scale strongly
and will differ from the metric in Planck units 
($g'_{\mu\nu}$) by a scale-dependent factor.
While this is not what one would ordinarily call a Quantum Gravity effect,
it is a quantum effect, and it affects our understanding
of the spacetime structure.
It will be interesting to see if this can have any
observable consequences.

\medskip
{\bf Acknowledgements.} I would like to thank S. Bertolini, M. Fabbrichesi,
L. Griguolo, D. Perini, M. Reuter, A. Romanino, E. Sezgin and A. Zaffaroni 
for discussions and comments.

\bigskip

{\bf Appendix A. Examples of essential and inessential couplings}

If a Lagrangian is parametrized in an arbitrary way, in general
no parameters can be straightforwardly absorbed into a field redefinition.
This is the case, for example, in a scalar field theory
with the following action:
$$
S=\int d^4x\left[
-{Z\over 2}\partial_\mu\phi\partial^\mu\phi
+\lambda_2\phi^2+\lambda_4\phi^4+\lambda_6\phi^6\ldots
\right]\ ,\eqno(A.1)
$$
None of the parameters appearing in this action satisfies the criterion 
for inessentiality given in [13], namely that
the variation of the Lagrangian with respect to the parameter
vanishes or is a total derivative on shell.
In fact, all parameters scale nontrivially under the $c$-transformations in (9).

If we redefine $\lambda_{2n}=Z^n\rho_{2n}$ the action becomes
$$
S=\int d^4x\left[
-{Z\over 2}\partial_\mu\phi\partial^\mu\phi
+Z\rho_2\phi^2+Z^2\rho_4\phi^4+Z^3\rho_6\phi^6\ldots
\right]\ \eqno(A.2)
$$
In this parametrization the couplings $\rho_{2n}$ are invariant under
the $c$-transformations, while $Z$ is not.
This is therefore an adapted parametrization of the action.
The wave-function renormalization is inessential
and indeed it can be absorbed by a simple rescaling
$\sqrt{Z}\phi=\psi$:
$$
S=\int d^4x\left[
-{1\over 2}\partial_\mu\psi\partial^\mu\psi
+\rho_2\psi^2+\rho_4\psi^4+\rho_6\psi^6\ldots
\right]\ \eqno(A.3)
$$
It is important to stress that this statement refers to the
parametrization (A.2): one cannot say that $Z$ in the action (A.1)
is inessential, because by this token one could conclude that {\it every}
parameter appearing in (A.1) is inessential.
For example, one could define
alternatively $\lambda_{2n}=\lambda_2^n\sigma_{2n}$ and $Z=\lambda_2 \zeta$,
in which case the action becomes
$$
S=\int d^4x\left[
-{\zeta\over 2}\lambda_2\partial_\mu\phi\partial^\mu\phi
+\lambda_2\phi^2+\lambda_2^2\sigma_4\phi^4+\lambda_2^3\sigma_6\phi^6\ldots
\right]\ ,\eqno(A.4)
$$
and in this parametrization it is the mass $\lambda_2$ that 
is inessential and disappears by the rescaling $\sqrt{\lambda_2}\phi=\Phi$. 

Similar considerations apply also to gravity, with the additional twist that
in this case eliminating an inessential parameter is equivalent 
to a choice of units of mass.
If we write a general action, somewhat symbolically, as
$$
S=\int d^4x\sqrt{g}\left[
a_0+a_1 R+a_2 R^2+a_3 R^3+\ldots
\right]\ \eqno(A.5)
$$
we can use the scale invariance with parameter $b$ in eq.(9)
to eliminate one of the couplings.
For example we can redefine $a_n=a_1^{2-n}a'_n$, ($a'_1=1$) leading to
$$
S=\int d^4x\sqrt{g}\left[
a_1^2 a'_0+a_1 R+a'_2 R^2+a_1^{-1}a'_3 R^3+\ldots
\right]\ \eqno(A.6)
$$
then rescaling $g_{\mu\nu}=a_1^{-1} g'_{\mu\nu}$, $a_1$ disappears:
$$
S=\int d^4x\sqrt{g'}\left[
a'_0+ R'+a'_2 R'^2+a'_3 R'^3+\ldots
\right]\ ,\eqno(A.7)
$$
which (up to trivial factors) is the action in Planck units.
So $a_1$ is inessential in the parametrization (A.6).
Alternatively we can
redefine $a_n=a_0^{{(2-n)\over2}}a''_n$, ($a''_0=1$), so that
$$
S=\int d^4x\sqrt{g}\left[
a_0+\sqrt{a_0}a''_1 R+a''_2 R^2+{a''_3\over\sqrt{a_0}} R^3+\ldots
\right]\ \eqno(A.8)
$$
Then defining $g_{\mu\nu}={1\over\sqrt{a_0}}g''_{\mu\nu}$, $a_0$ disappears
$$
S=\int d^4x\sqrt{g''}\left[
1+a''_1 R''+a''_2 R''^2+a''_3 R''^3+\ldots
\right]\ \eqno(A.9)
$$
This is the action written in units of the cosmological constant.
Note that one cannot eliminate $a_2$, which is invariant under $b$-transformations,
but one could eliminate $a_3$ or any one of the higher couplings.

\goodbreak
\bigskip 
\centerline{\bf References} 
\item{[1]} K. Wilson, Rev. Mod. Phys. {\bf 55}, 583 (1983); 
\item{[2]} M.J. Duff, L.B. Okun and G. Veneziano, JHEP 03, 023 (2002), {\tt physics/0110060}; 
F. Wilczek, Physics Today October 2005, January 2006, May 2006.
\item{[3]} P.A.M. Dirac, Proc. Roy. Soc. {\bf A 165}, 199 (1938); 
{\it ibid.} {\bf A 333}, 403 (1973); 
\item{[4]} J-P. Uzan,  Rev.Mod.Phys. {\bf 75}, 403 (2003), {\tt hep-ph/0205340}
\item{[5]} L. Randall and R. Sundrum, Phys. Rev. Lett. 83, 3370 (1999), {\tt hep-th/9905221}; 
\item{[6]} J. Maldacena, Adv. Theor. Math. Phys. 2, 231 (1998), {\tt hep-th/9711200};
E. Witten, Adv. Theor. Math. Phys. 2, 253 (1998), {\tt hep-th/9802150};
I.R. Klebanov, Introduction to the AdS/CFT Correspondence, 
Lectures at TASI '99, Boulder, June 1999; 
\item{[7]} H. Verlinde, Nucl. Phys. {\bf B 580} 264 (2000), {\tt hep-th/9906182}; 
N. Arkani-Hamed, M. Porrati and L. Randall, JHEP 0108:017 (2001), {\tt hep-th/0012148}; 
R. Rattazzi and A. Zaffaroni, JHEP 0104:021 (2001), {\tt hep-th/0012248};
\item{[8]} J. Polonyi and K. Sailer, Phys. Rev. D 63, 105006 (2001)
\item{[9]} H. Gies and C. Wetterich, Phys. Rev. D 65, 065001 (2002);
\item{[10]} J.M. Pawlowski, hep-th/0512261.
\item{[11]} R. Percacci and D. Perini, Class. and Quant. Grav. {\bf 21} 5035-5041 (2004) {\tt hep-th/0401071};
\item{[12]} H.W. Hamber, Phys.Rev. D 61 124008 (2000), {\tt hep-th/9912246};
\item{[13]} S. Weinberg, in ``General Relativity: An Einstein centenary survey'', 
edited by S.~W. Hawking and W. Israel, Cambridge University Press, 1979, chapter 16, pp.790--831; 
\item{[14]} M. Reuter, Phys. Rev.{\bf D57}, 971 {1998}, {\tt hep-th/9605030}; 
\item{[15]} D. Dou and R. Percacci, Class. Quant. Grav. {\bf 15}, 3449, (1998), {\tt hep-th/9707239}; 
\item{[16]} W. Souma, Prog. Theor. Phys. {\bf 102}, 181 (1999), {\tt hep-th/9907027}; 
O. Lauscher and M. Reuter, Phys. Rev. {\bf D65}, 025013 (2002), {\tt hep-th/0108040}; 
O. Lauscher and M. Reuter, Class. Quant. Grav. {\bf 19}, 483 (2002), {\tt hep-th/0110021}; 
O. Lauscher and M. Reuter, Int. J. Mod. Phys. {\bf A17}, 993 (2002), {\tt hep-th/0112089}; 
O. Lauscher and M. Reuter, Phys. Rev. {\bf D66}, 025026, (2002), {\tt hep-th/0205062}; 
\item{[17]} R. Percacci and D. Perini, Phys. Rev. {\bf D67}, 081503 (2002), {\tt hep-th/0207033}; 
R. Percacci and D. Perini, Phys. Rev. {\bf D68}, 044018 (2003), {\tt hep-th/0304222}; 
\item{[18]} O. Lauscher and M. Reuter, {\it Phys. Rev.} {\bf D 66}, 025026;
R. Percacci, Phys.Rev. {\bf D73}:041501 (2006), {\tt  hep-th/0511177};
A. Codello and R. Percacci, Phys. Rev. Lett. 97, 221301 (2006) {\tt   hep-th/0607128};
\item{[19]} L. Griguolo and R. Percacci, Phys.Rev. {\bf D 52}, 5787, (1995), 
{\tt hep-th/9504092}; 
\item{[20]} E. Tomboulis, Phys. Lett. {\bf B 389}, 225 (1996), {\tt hep-th/9601082};
\item{[21]} J. Magueijo and L. Smolin, Phys.Rev. {\bf D67} 044017 (2003), {\tt gr-qc/0207085}.
\item{[22]} F. Wilczek, Nature {\bf 397}, 303 (1999); Physics Today, November 1999, June 2001, November 2001, August 2002;
\item{[23]} S. Bornholdt, C. Wetterich, Phys.Lett. {\bf B282} 399-405 (1992);
\item{[24]} C. Gomez, B. Janssen and P. Silva, JHEP {\bf 0004} 024 (2000), {\tt hep-th/0002042};
Phys.Rev. {\bf D 65}, 043508, (2002), {\tt hep-th/0106133};
\item{[25]} A. Bonanno and M. Reuter, Phys.Rev. {\bf D 65}, 043508, (2002), {\tt hep-th/0106133};
\item{[26]} R. Bousso, in ``Cargese 2002, Progress in string, field and particle theory'',
proceedings of the NATO Advanced Study Institute and EC Summer School on Progress in String, Field and Particle Theory, 
Cargese, Corsica, France, 25 Jun - 11 Jul 2002, {\tt hep-th/0203101}.

\vfil
\eject
\bye